# ResearchGate and Google Scholar: How much do they differ in publications, citations and different metrics and why?

**Vivek Kumar Singh[1], Satya Swarup Srichandan & Hiran H. Lathabai**

Department of Computer Science, Banaras Hindu University, Varanasi-221005, India.

**Abstract:** ResearchGate has emerged as a popular professional network for scientists and researchers in a very short span of time. Similar to Google Scholar, the ResearchGate indexing uses an automatic crawling algorithm that extracts bibliographic data, citations and other information about scholarly articles from various sources. However, it has been observed that the two platforms often show different publication and citation data for the same institutions, journals and authors. This paper, therefore, attempts to analyse and measure the differences in publications, citations and different metrics of the two platforms for a large data set of highly cited authors. The results indicate that there are significantly high differences in publications and citations for the same authors captured by the two platforms, with Google Scholar having higher counts for a vast majority of the cases. The different metrics computed by the two platforms also differ in their values, showing different degrees of correlations. The coverage policy, indexing errors, author attribution mechanism and strategy to deal with predatory publishing are found to be the main probable reasons for the differences in the two platforms.

**Keywords:** Academic Social Networks, Altmetrics, Google Scholar, ResearchGate, Scientometrics.

**Introduction**

ResearchGate[2], an academic social networking site created in 2008, has emerged as a popular professional network for scientists and researchers in a very short span of time. It had only 10,000 members in 2008, but has grown to more than 20 million members at present[3]. ResearchGate states that its mission is "*to connect the world of science and make research open to all*". It provides a number of features that allows its members to share and discover research. The members can create a profile and upload their research papers and projects, can ask and answer questions, can create projects and collaborate on that with other members, and get notified about a number of events ranging from reads to citations. Several previous studies have explored the repository feature of ResearchGate (such as Borrego, 2017; Jamali, 2017; Van Noorden, 2017), while some others examined the various aspects of usage patterns of ResearchGate site (such as Ortega, 2017; Muscanell & Utz, 2017; Yan & Zhang, 2018; Meier & Tunger, 2018; Lee et al., 2019; Mason & Sakurai, 2020; Ebrahimzadeh et al., 2020; Yan et al., 2021).

Google Scholar[4] is a popular freely accessible search engine in the academic domain. It was created in 2004 and indexes full-text or metadata of scholarly literature. It has a very wide coverage including most of the peer-reviewed online academic journals and books, conference papers, theses and dissertations, preprints, abstracts, technical reports, and other scholarly literature[5]. Scientometric studies have indicated that it is the world's largest academic search engine, indexing roughly 389 million documents including articles, citations and patents as in 2018 (Gusenbauer, 2019). However, Google Scholar have often been criticized on account of its accuracy and due to including predatory journals in its index (Beall, 2014; Kolata, 207). Google Scholar also tracks citations to articles it indexes and in the year 2012, it added the feature of individual scholars to create personal Scholar Citations Profiles[6]. It now publishes a number of metrics for scholars including total citations, h-index and i-10 index.

---

[1] Corresponding author. Email: vivek@bhu.ac.in
[2] htttps://www.researchgate.net/
[3] https://www.researchgate.net/press
[4] http://scholar.google.com

ResearchGate also computes and publishes a number of metrics (such as reads, citations etc.) and also assigns an RG score to its members, though it is not fully known how exactly this score is computed from different components of a member's profile. Many studies analysed the different metrics of ResearchGate, with respect to their correlations with metrics from other bibliometric databases, rankings and online platforms (Thelwall & Kousha, 2015; Kraker & Lex, 2015; Jordan, 2015; Shrivastava & Mahajan, 2015; Nicholas, Clark & Herman, 2016; Yu et al., 2016; Orduna-Malea et al., 2017; Thelwall & Kousha, 2017a, 2017b; Shrivastava & Mahajan, 2017; Lepori, Thelwall & Hoorani, 2018; Copiello & Bonifaci, 2018; Copiello & Bonifaci, 2019; Copiello, 2019; Banshal, Singh & Muhuri, 2021). The RG score has been explored quite a lot in several respects, including its computation & reproducibility (Kraker & Lex, 2015; Jordan, 2015, Copiello & Bonifaci, 2018; Copiello & Bonifaci, 2019; Copiello, 2019), its correlation with related measures from bibliometric databases (Thelwall & Kousha, 2015; Shrivastava & Mahajan, 2015; Thelwall & Kousha, 2017a, 2017b), and its usefulness as a measure of academic reputation (Yu et al., 2016; Orduna-Malea et al., 2017; Lepori, Thelwall & Hoorani, 2018).

Despite many previous studies of different kinds on both Google Scholar and ResearchGate, the two platforms have not been explored much together. Thelwall & Kousha (2017b) and Ortega (2017) are perhaps the only among many studies on ResearchGate, that explored ResearchGate and Google Scholar together, though in different respects. Thelwall & Kousha (2017b) analysed citations recorded by ResearchGate and compared it with the values of Google Scholar, at the level of journals. They observed that ResearchGate found statistically significantly fewer citations than did Google Scholar. It was suggested that ResearchGate and Google Scholar may be predominantly tapping similar sources since ResearchGate citations correlated strongly with Google Scholar citations. However, this study limited itself only to the examination of the citations and other metrics maintained and computed by ResearchGate (such as RG score, reads etc.) were not compared with metrics recorded by Google Scholar (such as h-index and i-10 index). Ortega (2017) explored the presence and disciplinary variations of profiles of one organization (Consejo Superior de Investigaciones Científicas (CSIC)) in three major academic social sites (Academia.edu, Google Scholar and ResearchGate). However, the study was limited only to the analysis of similarities in presence of researchers and did not compare the different metric values in the platforms.

Given the fact that both ResearchGate and Google Scholar use an automated Web crawling approach for data collection, it would be very important to explore if the bibliographic data (publications and citations) and metrics (h-index, RG score etc.) in the two platforms, for the same set of authors, are similar or different. Further, if these counts and metric values differ, what is the quantum of difference and what could be the possible reasons for such differences. While some previous studies (Orduna-Malea & López-Cózar, 2017; Martín-Martín, Orduña-

---

[5] https://scholar.google.com/intl/us/scholar/help.html#coverage
[6] https://scholar.googleblog.com/2014/08/fresh-look-of-scholar-profiles.html

Malea & López-Cózar, 2018) analysed different Author Level Metrics (ALM) for several platforms, including ResearchGate and Google Scholar; but they had a different focus. These studies limited themselves to observe correlations in the ALM values in ResearchGate and Google Scholar and did not dwell into measuring the difference in values of different counts and metrics and explore reasons thereof. This article, therefore, attempts to address this gap in knowledge by computing the difference in the bibliographic data (publications and citations) and metric values (h-index, RG score etc.) in the two platforms for a large set of highly cited authors. The analysis includes publications, citation counts, reads, h-index, estimated h type indices and the RG score values.

The rest of the article is organized as follows: The section on 'Related Work' presents a brief survey of some of the most relevant studies on ResearchGate platform and its relatedness with other academic social networks and scholarly databases. The 'Data and Methodology' section presents details of the data used for the analysis and the computational approach used for the analysis. The 'Results' section describe the various analytical results obtained about the differences in bibliographic data and metric values and the different types of correlations observed. The 'Discussion' section briefly discusses the observed results along with probable reasons for such patterns. The paper concludes with the main conclusions drawn from the study in the 'Conclusion' section.

**Related Work**

ResearchGate, being an academic social network, has a mix of features of social networks and bibliographic databases. The increasing popularity of ResearchGate has attracted attention of researchers, who explored its various features. It has been explored both as a repository of scientific publications and as a platform for information sharing and dissemination. The different metrics computed by ResearchGate, with RG score being the most notable one, have also been explored in several studies. This section presents a survey of some of the important studies on ResearchGate and its relatedness with other academic social networks and scholarly databases.

*Studies on ResearchGate as a Repository*

Borrego (2017) compared the availability of the scholarly output of some top Spanish universities in their institutional repositories and in ResearchGate (RG). Data for 13 Spanish universities ranked among the top 500 universities worldwide in the 2015 ARWU rankings was obtained from Web of Science and analysed. For the total 30,630 articles retrieved, availability in the two sources: the university's IR and RG were examined. It was found that only 11% of the articles published by scholars based at top Spanish universities are available in their institutional repository. However, more than half of the articles published by these scholars are available in full text in ResearchGate. It was found that many researchers were unaware of institutional repositories but were appreciative of the advantages offered by academic social networking sites such as ResearchGate.

Muscanell & Utz (2017) examined the usage and utility of ResearchGate (RG) by analysing the ways different authors use the site, their views on site's utility, and the effects of usage on their career outcomes. An online survey approach was used and data was collected at three time points; during the winter of 2014, summer of 2014 and spring of 2015. A total of 1009 participants in age from 18 to 86 (mean age = 39) from United States (60.1%) and Europe (32.9%) were targeted. The results have shown that most academics who have an RG account did not use it very heavily. Further, users did not perceive many benefits from using the site, and RG use was not related to career satisfaction or informational benefits, but was related to productivity and stress. The study concluded that systematic research is needed to explore positive and negative consequences of using professional social media in academia, especially productivity and stress. The findings also suggested that RG needs to increase user engagement.

Jamali (2017) investigated that to what extent the ResearchGate members, as authors of journal articles, comply with publishers' copyright policies when they self-archive full-text of their articles on ResearchGate. A random sample of 500 English journal articles available as full-text on ResearchGate were investigated. It was found that 108 articles (21.6%) were open access (OA) published in OA journals or hybrid journals. Out of the remaining 392 articles, 61 (15.6%) were preprint, 24 (6.1%) were post-print and 307 (78.3%) were published (publisher) PDF. It was observed that 201 (51.3%) out of 392 non-OA articles infringed the copyright and were non-compliant with publishers' policy. While 88.3% of journals allowed some form of self-archiving (Sherpa Romeo green, blue or yellow journals), the majority of non-compliant cases (97.5%) occurred when authors self-archived publishers' PDF files (final published version). This indicated that authors infringe copyright most of the time not because they are not allowed to self-archive, but because they use the wrong version, which might imply their lack of understanding of copyright policies and/or complexity and diversity of policies.

Van Noorden (2017) examined the copyright infringement issues in uploads to ResearchGate site. He suggested that millions of articles might soon disappear from ResearchGate due to issues related to breach of publishers' copyright. Some publishers formed a coalition and filed a lawsuit to try to prevent copyrighted material appearing on ResearchGate in future. The complaint was filed in a regional court in Germany as ResearchGate is based in Berlin. Though the complaint made a "symbolic request for damages" but apparently its main goal was to change the site's behaviour. In fact, ResearchGate removed "a significant number of copyrighted articles". It was observed that not only do academics upload articles to the site, but ResearchGate also scrapes material online and invites researchers to claim and upload these papers.

Lee et al. (2019) investigated the motivations for self-archiving research items on academic social networking sites (ASNSs), mainly ResearchGate. A model of these motivations was developed based on two existing motivation models: motivation for self-archiving in academia and motivations for information sharing in social media. The proposed model is composed of 18 different factors. Data of survey response from 226 ResearchGate users was analysed. It was observed that accessibility was the most highly rated factor, followed by altruism, reciprocity, trust, self-efficacy, reputation, publicity, and others. Personal, social, and professional factors were also highly rated, while external factors were rated relatively low. They concluded that self-archiving research on RG is a way to promote open science that makes scientific results accessible to and reusable by a wider audience; which in turn creates an era of networked science, thus accelerating the progress of science.

### *Studies on usage patterns and social network features of ResearchGate*

Ortega (2017) analysed the distribution of profiles from academic social networking sites according to disciplines, academic statuses and gender, and detect possible biases. The profiles of one organization (Consejo Superior de Investigaciones Científicas (CSIC)) in three major academic social sites (Academia.edu, Google Scholar Citations and ResearchGate) through six quarterly samples (April 2014 to September 2015) were tracked. A total of 7,193 profiles were retrieved belonging to 6,206 authors. It was found that most of the CSIC profiles were there on ResearchGate (4,001 profiles) as compared to Google Scholar (2,036 profiles) and

Academia.edu (1,156 profiles). Results have also shown disciplinary bias but gender distribution does not display strong differences.

Yan & Zhang (2018) examined the institutional differences, research activity level, and social networks formed by research universities on the ResearchGate (RG) social networking site. The study collected data from RG users from 61 U.S. research universities at different research activity levels, as categorized by the Carnegie Classification of Institutions of Higher Education. The purpose was to examine the impact of institutional differences on RG reputational metrics. The data was crawled during 22nd September to 2nd October, 2016, with 168,059 users' data collected. For each user in the sample, the top 10 followers and top 10 followees' institution data were collected and combined according to institution. Data for 459,763 followers and 360,250 followees was obtained. The analysis showed that with an increase in the research activity level of a university, its affiliated RG users tend to have higher RG scores, more publications and citations, and more profile views and followers. The study suggested that academic social networks can serve as indicators in evaluation of research activities among research institutions.

Meier & Tunger (2018) investigated opinions and usage patterns relating to the ResearchGate social networking site for scientists and researchers. They used a survey that consisted of 19 questions and was conducted online with 695 scientists from the disciplines of physics, biology, medicine, and neuroscience. The research questions concerned how much time and effort the interviewees expended on ResearchGate, what added value they perceived in using the site, the extent to which social aspects influence use, how participants planned to use the platform in future, and what role ResearchGate's own metric, the RG score, played for the scientists. In addition, factors of age, sex, origin, and scientific discipline have been explored. The survey responses revealed that the respondents invest relatively little time in browsing ResearchGate or updating their own profiles. Free access to publications was the most frequently mentioned benefit by the respondents along with the opportunity to exchange ideas with other scientists. To conclude, majority of participants are of the opinion that it makes sense for scientists to use ResearchGate.

Mason & Sakurai (2020) performed a national study of Japan through a survey of the use of RG by more than 500 researchers across the country. The authors investigate participants' awareness and regularity of use of the 3 micro-level components of RG, and the benefits and challenges of their adoption. They observed that RG is largely perceived as valuable for participants but use is unbalanced toward knowledge sharing. It was found that one of the major uses of RG for participants is access to knowledge, particularly through posted research outputs. Participants also use RG as a platform to disseminate their own research, which is noted as a major benefit. One of the major barriers to RG use is in the area of knowledge sharing, and relates to copyright issues. They concluded that RG may be positioned as a tool rather than a community.

Ebrahimzadeh et al. (2020) attempted to identify the triggers, strategies and outcomes of collaborative information-seeking behaviours of researchers on the ResearchGate social networking site. They analysed qualitative interview data for the Ph.D. students and assistant professors in the library and information science domain. They observed that informal communications and complex information needs lead to a decision to use collaborative information-seeking behaviour. Further, easy access to sources of information and finding relevant information were found to be the other major positive factors contributing to collaborative information-seeking behaviour of the ResearchGate users.

Yan et al. (2021) explored how the scholarly use of academic social networking sites (mainly ResearchGate) differ by academic discipline. The study collected data from a total of 77,902 users from 61 U.S. research universities at different research activity levels as defined by the Carnegie Classification of Institutions of Higher Education. The disciplinary comparison of this study showed significant differences in users' participation and utilization of RG between disciplines. Life Sciences & Biomedicine users embraced RG the most and boasted notable academic influence, which is reflected in various RG metrics. Physical Sciences users positively updated publications, and Social Sciences users concentrated more on networking and maintaining creditable reputations. In contrast, Technology users demonstrated moderate participation levels with less recognized efforts, and Arts & Humanities users exhibited an overall lower utilization of RG. In addition, users from higher research activity level universities tend to show better performance in RG metrics than their lower research activity level counterparts regardless of discipline. Thus, the study concluded that the user participation and RG use characteristics vary by discipline.

*Studies on RG Metrics and their correlation with other bibliographic databases*

Thelwall & Kousha (2015) explored the dissemination, communication and measurement of scholarship in the ResearchGate platform. The objective was to find out whether ResearchGate usage broadly reflects the traditional distribution of academic capital and to what extent its metrics correlate with traditional academic rankings at the university level. They obtained list of institutional homepages in ResearchGate as in October 2013 (over 31,000 URLs) and the publication statistics from Web of Science. The comparisons of the rankings between institutions found that total impact points correlated less with the different rankings than they did with each other, though the difference was not large. They concluded that ResearchGate is changing scholarly communication and that ResearchGate view counts and download counts for individual articles may prove to be useful indicators of article impact in the future.

Kraker & Lex (2015) presented an assessment of the ResearchGate score as a measure of a researcher's scientific reputation. This assessment was based on well-established bibliometric guidelines for research metrics. It was observed that the ResearchGate Score has three serious shortcomings: (1) the score is no transparent and irreproducible, (2) the score incorporates the journal impact factor to evaluate individual researchers, and (3) changes in the score cannot be reconstructed. They concluded that ResearchGate Score should not be considered in the evaluation of academics in its current form. Jordan (2015) presented a response to the abovementioned work of Kraker & Lex (2015). They undertook a small-scale exploratory analysis of ResearchGate scores to examine correlations between ResearchGate score and profile metrics. The importance of the Journal Impact Factor in determining ResearchGate score was confirmed. They found that RG score offers advantages over the JIF and citation counts, as it has the potential to account for alternative ways of measuring activity and impact. They noted that the RG score claims to intend to do so via academics' social interactions on the site. However, they also observed that there is a mismatch between the goal of the RG score and use of the site in practice, which may amplify the influence of the JIF upon RG score. They concluded that most academics who use ResearchGate view it as an online business card or curriculum vitae rather than a site for active interaction with others.

Shrivastava & Mahajan (2015) investigated the relationship between the altmetric indicators from ResearchGate (RG) and the bibliometric indicators from the Scopus database. They collected data manually by visiting the profile pages of all the members of the Department of Physics, Panjab University, Chandigarh (India), having a ResearchGate account, during the first week of December 2014. The data for a total of 70 members was collected from ResearchGate, including publications, profile views, publication views, citations, impact points and RG Score. They found that most of the RG metrics showed strong positive correlation with the Scopus metrics, except for RG Score (RG) and Citations (Scopus), which showed moderate positive correlation. It was also found that the RG metrics showed moderate to strong positive correlation amongst each other.

Nicholas, Clark & Herman (2016) explored the ResearchGate (RG)'s reputational facilities, which involved identifying, explaining, evaluating, testing and comparing all its mechanisms and metrics. They collected RG profile of 400 researchers and investigated their profiles and scores. They observed that RG offers a great deal when it comes to building, showcasing, and measuring reputation. The RG Score acknowledges the fact that reputation is, in its essence, social and collaborative; however, it struggles with the deployment of engagement metrics. They concluded that RG has the potential to upset the reputational applecart by becoming a major deliverer of scholarly reputation.

Yu et al. (2016) compared the ResearchGate metrics with the Research Excellence Framework (REF) and Quacquarelli Symonds (QS) World University Rankings to assess the quality of UK universities and global universities, respectively. The study used correlation analysis to examine whether ResearchGate metrics demonstrate effectiveness on the researcher level in comparison with SciVal metrics. They analysed data for 300 ResearchGate members from the field of supply chain management. They found that ResearchGate score exhibits a moderate correlation with REF metrics, it exhibits a strong correlation with selected QS metrics. The analytical results provided empirical evidence that the ResearchGate score can be an effective indicator for measuring individual researcher performance.

Orduna-Malea et al. (2017) investigated whether it is reasonable to employ the RG Score as evidence of scholarly reputation. They used three data samples: (a) an outlier sample of 104 authors with high values, (b) a Nobel sample comprising of 73 Nobel winners from Medicine and Physiology, Chemistry, Physics and Economics (from 1975 to 2015), and (c) a longitudinal sample of weekly data on 4 authors with different RG Scores. The analytical results suggest showed that RG Scores are built mainly from activity related to asking and answering questions in the site. They observed that Answers dimension is more influential than the remaining categories (Publications, Questions, and Followers). Active participation through questions, though important, seems to be less influential. The relationship between publications and RG Score were confirmed to be logarithmic, making it difficult to achieve a high score from publications alone. The results pointed to the existence of two different worlds within prominent ResearchGate members. The first (academics) is constituted from authors with many scientific publications and high bibliometric indicators (productivity, citation, and h-index). The second (active RG users) is formed from authors who build their reputation through their communication and collaboration activities within the site. They concluded that the RG Scores should not be mistaken for academic reputation indicators.

Orduna-Malea & Delgado López-Cózar (2017) studied the performance behavior patterns in Author-Level Metrics (ALM) from Google Scholar Citations, ResearchGate and ImpactStory. They analysed two kinds of author samples (intra and inter) and observed a non-linear distribution in the ALM data extracted from the three platforms (Google Scholar Citations, ResearchGate, and ImpactStory). They found that there are few authors with a high performance, and a long tail with moderate, low, or null performance. However, the high-performance authors are not the same across the three studied dimensions of impact (Citations, Reads, and Online mentions). They concluded that lack of correlation in ALM from the three platforms might be explained by the fact that each platform offers different documents, targeted to different audiences.

Thelwall & Kousha (2017a) explored the age and discipline of the articles uploaded and viewed in the ResearchGate site and that whether publication statistics from the site could be useful impact indicators. They assessed samples of ResearchGate articles uploaded at specific dates (Jan. 2014, July 2014 and Jan. 2015) and compared their views in the site to their Mendeley readers and Scopus-indexed citations. They found that ResearchGate is dominated by recent articles, which attract about three times as many views as the older articles. Further, ResearchGate was found to have an uneven coverage of scholarship, with the arts and humanities, health professions, and decision sciences poorly represented as compared to other disciplines. Some fields were found to receive twice as many views per article as compared to other fields. The analysis found that view counts for uploaded articles have low to moderate positive correlations with both Scopus citations and Mendeley readers.

Thelwall & Kousha (2017b) analysed that which out of ResearchGate, Google Scholar, WoS and Scopus gives the most citations for recently published library and information science journal articles. Further, they examined the similarity of the rank orders of articles produced by the different sources. The study used data for English language research or review articles published in 86 Information Science and Library Science (IS & LS) journals during January 2016 to March 2017 from Web of Science (WoS). They observed that ResearchGate found statistically significantly fewer citations than did Google Scholar, but more than both Scopus and Web of Science. Google Scholar always showed more citations for each individual journal than ResearchGate, though ResearchGate showed more citations than both WoS and Scopus. It was found that ResearchGate correlated strongly with Google Scholar citations, suggesting that ResearchGate is not predominantly tapping a fundamentally different source of data than Google Scholar.

Shrivastava & Mahajan (2017) performed an altmetric analysis of RG profile and RG score of faculty members from a department in a selected University. The data were collected manually in July 2016 by visiting the profile pages of all the members who had an account in ResearchGate under University of Delhi in India. A total of 173 members were found in ResearchGate from the department. Data were collected for publications, reads, profile views, citations, impact points, RG Score, followers and following from the profile pages of the members. Correlations were calculated amongst the metrics provided by ResearchGate to seek the nature of the relationship amongst the various ResearchGate metrics. The analysis has shown that publications added by researchers to their profiles were relatively low. The highest correlation of RG Score was found with publications added by researchers to their profiles. This was followed by correlation between RG Score and reads, correlation between RG score and profile views, correlation between RG Score and number of Full Texts and correlation between RG Score and number of followers of a researcher, in the order.

Martín-Martín, Orduña-Malea & López-Cózar (2018) analysed the Author Level Metrics (ALM) in the new academic profile platforms (Google Scholar Citations, ResearcherID, ResearchGate, Mendeley and Twitter). Data for a sample of 811 authors in the field of Bibliometrics was used to analyse a total of 31 ALMs. Two kinds of ALMs were identified, first related to connectivity (followers), and second to academic impact. The second group was further divided into usage metrics (reads, views), and citation metrics. They observed that Google Scholar Citations provides more comprehensive citation-related data as compared to other platforms. Twitter stands out in connectivity-related metrics. ResearchGate also showed a significant amount of social interaction among researchers (followers/ followees), much higher than in Mendeley.

Lepori, Thelwall & Hoorani (2018) explored about the visibility of US and European higher education institutions in ResearchGate (RG) platform and analysed what factors affect their RG score. It was found that most of the 2,258 European and 4,355 US higher educational institutions included in the sample had an institutional ResearchGate profile, with near universal coverage for PhD-awarding institutions found in the Web of Science (WoS). For non-PhD awarding institutions that did not publish, size (number of staff members) was most associated with presence in ResearchGate. For PhD-awarding institutions in WoS, presence in RG was strongly related to the number of WoS publications. It was concluded that: (a) institutional RG scores reflect research volume more than visibility (a bit of contrast to observations of Orduna-Malea et al. (2017)), and that (b) this indicator is highly correlated to the number of WoS publications. Therefore, the usefulness of RG Scores for institutional comparisons is limited.

Copiello & Bonifaci (2018) evaluated ResearchGate score as an indicator of scientific and academic reputation. They explored what does the expression ''academic reputation'' mean and how can it be measured in reference to ResearchGate platform. They used data of 26 authors with an RG Score higher than 100, and the 25 Nobel authors with the highest RG Score from the work of Orduna-Malea et al. (2017). They analysed RG Score, number of publications (including only articles, books, chapters, and conference papers), and number of full-text sources (regardless of the contribution item they refer to) of the authors. Through the analysis they concluded that RG Score is not a reliable indicator of scientific and academic reputation rather its more like a tool to implement the entrepreneurial strategy of the RG's owner company. Copiello & Bonifaci (2019) did a follow up study on ResearchGate Score, full-text research items, and full-text reads. They examined the relationship among full-text research items uploaded in the platform, full-text reads of the same items, and the RG Score. For this purpose, they analysed the RG Score, number of research items, full-texts, overall reads, and full-text reads for two data samples. Three ratios: full-texts to research items (F-t/I), full-text reads to overall reads (F-tr/R), and full-text reads to full texts (F-tr/F-t) were computed. They found that the ratio between full-text reads and full-texts shows a less significant difference between the samples. The ratio between full-texts and research items positively correlates with overall reads and, in turn, both the previous parameters positively correlate with the RG Score. However, it was found that RG Score is independent of full-texts and full-text reads. Thus, they concluded the paper with the question "are we really willing to let an (unknown) algorithm decides who and what deserves recognition in science?" In another study Copiello (2019) explored the Research Interest (RI) algorithm of ResearchGate to see if it is truly transparent and whether it is fully disclosed. They used part of data explored by Orduna-Malea et al. (2017) and observed that RI score is built upon a weighting scheme of citations, recommendations, full-text reads, and other reads by RG members. The study concluded that the RI indicator suffers from, at least, two significant issues: lack of transparency and redundancy.

Banshal, Singh & Muhuri (2021) explored whether early altmetric mentions (including RG reads) could predict future citations, by identifying the type and degree of correlations between altmetrics and citation for a large sample of scholarly articles taken for India. They observed that altmetrics do not necessarily reflect the same kind of impact as citations. However, articles that get higher altmetric attention early may actually have a slight citation advantage. Further, altmetrics from academic social networks like ResearchGate are more correlated with citations, as compared to altmetric events of social media platforms.

**Data & Methodology**

The data for analysis comprises of publication and citation data, along with some other metrics, that are obtained from Google Scholar and ResearchGate. First of all, the list of authors having h-index value greater than 100, in Google Scholar as in the month of Feb. 2021, is obtained. There is a total of 4,238 such authors. The next step was to find out which of these authors have profiles in ResearchGate. For this purpose, we designed a customized crawler to automatically query the ResearchGate website. Thus, for each author in the list of 4,238 authors with Google Scholar profile, the ResearchGate was crawled for a matching profile. For ensuring that the ResearchGate profile retrieved is for the same author as in Google Scholar, the ResearchGate and Google Scholar profile for each author were reproduced in side-by-side windows and information such as name, place, university, profile picture etc. in the two windows were compared manually. A total of 1,758 authors were found to have profiles in both- Google Scholar and ResearchGate. Now, for each of these 1,758 authors, their profiles in Google Scholar and ResearchGate were downloaded and saved locally. The Google Scholar profile of authors contained information about the publications, citations, h-index, i-10 index, current university and position, current place (city & country) etc. The ResearchGate profile of authors contained information about publications, reads, citations, RG Score, current university and position, current place, academic degree etc. For few authors, there were two ResearchGate profiles and therefore we selected the updated one with more data.

Out of the different kinds of information available in the Google Scholar and ResearchGate profiles of the 1,758 authors (common in both the platforms), we first analysed the publications and citations for each author in the two platforms. The difference in publication and citations was computed for each author. Thereafter, the h-index values from the Google Scholar and ResearchGate was compared for all the authors. Since the two platforms only provide h-index values, we used the publication-citation data to estimate values of other h-type indices and observe if they correlate for different kinds of profiles: regular citation profile, citation excess profile and transition profiles, as described further in the Results section. The RG Score from ResearchGate was also compared with Google Scholar h-index. The values of other counts and metrics (such as reads) for the same authors in the two platforms were also compared. Rank correlations between different metrics for the authors in the two platforms are also computed by calculating Spearman Rank Correlation Coefficients.

**Results**

The data from Google Scholar (GS) and ResearchGate (RG) has been analysed to measure differences in publications, citations and the various metrics in the two platforms. The subsections below describe the observed patterns.

*Number of publications in GS and RG*

First, we analysed the publication data in the two platforms for all the authors. **Figure 1** shows the publication counts for all the 1,758 authors in GS and RG. Here, the y-axis represents the number of publications and the x-axis represent authors and the author names are arranged in the decreasing order of their GS h-index. It is observed that GS has higher number of publications recorded for majority of the authors. To illustrate further, 1,597 out of 1,758 authors (i.e. 90.8%) record higher publication counts in GS, and only 161 out of 1,758 authors record higher publication counts in RG. The differences are also quite high. For example, the author 'Ronald C. Kessler' who has 1,910 publications in GS has only 1,080 publications in RG (thus GS having 830 more papers). However, there are counter examples as well. For example, author 'Rob Knight' has 847 papers in GS but 1,282 papers in RG.

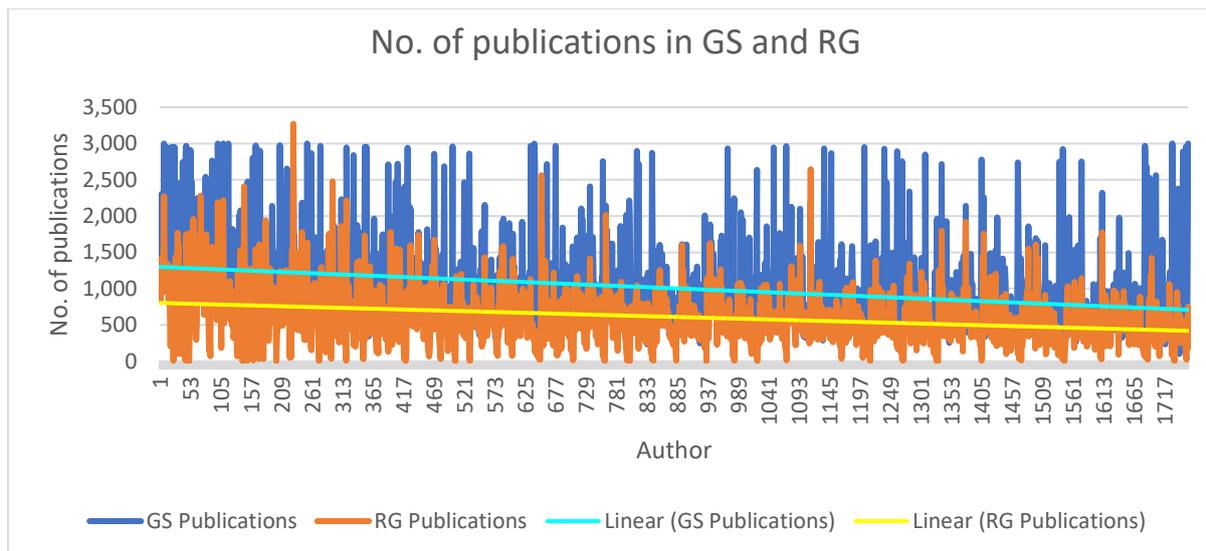

**Figure 1: Publication counts of authors in GS and RG**

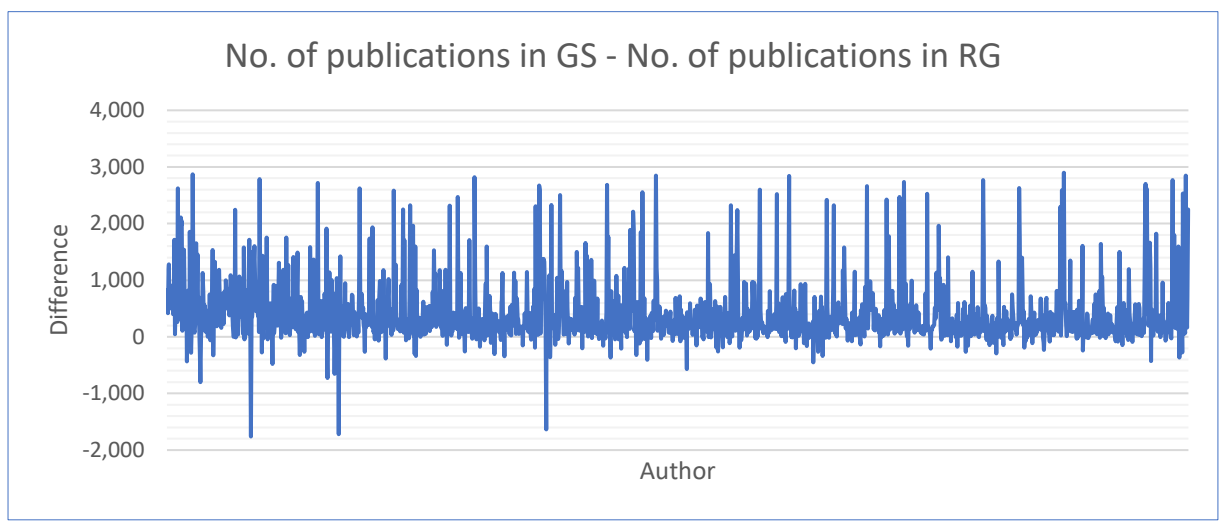

**Figure 2: Difference in publication counts of authors in GS and RG**

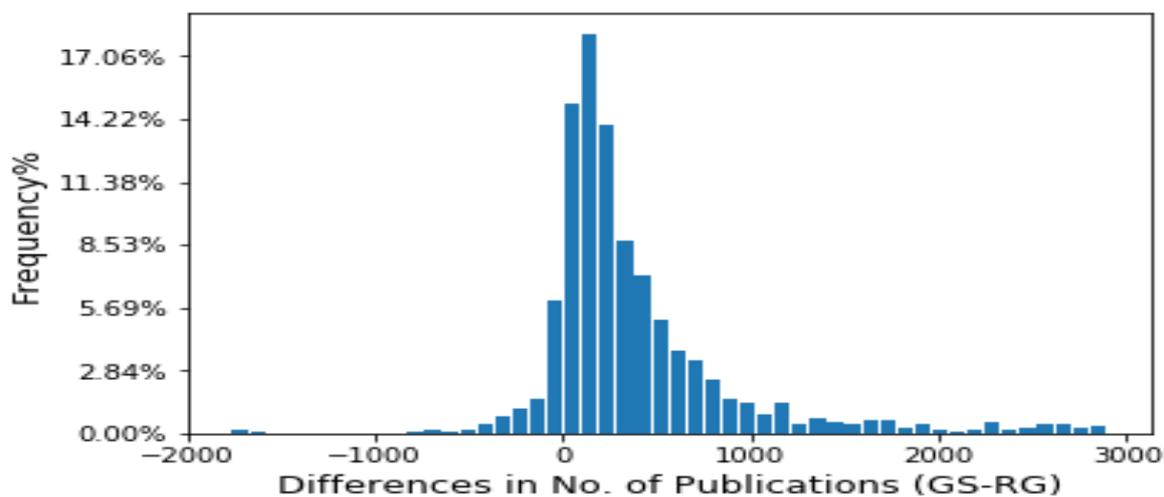

**Figure 3: Frequency of difference in publication counts of authors in GS and RG**

The difference in publication counts for the same authors in GS and RG is observed to be significantly high. **Figure 2** plots the difference of No. of publications in GS and No. of publications in RG for all the authors. The difference is positive for 1,597 authors and negative for 161 authors. The mean of difference in publication counts is 394. Thus, it is quite clear that for the given set of authors, GS records significantly greater number of publications than RG. The variance is 281,944 and standard deviation is 531. **Figure 3** shows more clearly the frequency of differences in publication counts in different intervals. As can be seen in the figure, the difference values are inclined towards the right side of the '0' value indicating a large percentage of authors having the difference value as positive. It can be observed that about 75% authors have the difference in GS and RG publications in the range 0-1,000. About 10% authors have the difference in GS and RG publications in the range of 1,000-2,000. For about 3-4% of authors, this difference is more than 2,000+ publications. Thus, not only GS records higher number of publications than RG for majority of authors but the magnitude of difference is also significantly high.

*Number of citations in GS and RG*

The next parameter analysed was the total number of citations for all the authors as recorded by the two platforms. As observed in the section above that for the majority of the authors in the given set, GS records higher publications than RG, therefore, one may expect that total citation counts for authors recorded by GS may be more than RG. **Figure 4** plots the total citation counts for all the authors in GS and RG platforms. Here, the y-axis represents the total number of citations and x-axis represents the authors. It is observed that for a significantly large number of authors, GS records higher number of total citations as compared to RG. In fact, for 1,753 out of 1,758 authors (i.e. 99.7%), GS records higher citation counts as compared to RG. There are only 5 authors for whom RG records more citations than GS. Though there are 161 authors having more publications recorded in RG, but only 5 authors have higher citation counts recorded in RG.

**Figure 5** shows the difference of No. of citation counts in GS and No. of citations counts in RG for all the authors. It can be seen that almost the entire plot is above the difference=0 line. Further, the differences are high, with a mean value of 34,937 citations. The variance value is 1,215,294,779 and the standard deviation is 34,861. For a clearer visualization, **Figure 6** shows the frequency of citation difference in different intervals. It can be seen that except few, most of the difference values are on the positive side of '0' value. The difference in citations is significantly large, as we see that there are some authors with the difference value more than even 200,000. We see that for about 70% authors, the difference in citations is in the range of 0 - 50,000. There are also a good number of authors who have the difference in citations more than 50,000. Thus, GS is observed to record significantly large number of citations as compared to RG.

There are many interesting examples to look at. For many authors who have higher number of publications in RG, the number of total citations recorded by RG is lower than GS. For example, author 'Rob Knight' who has 435 more publications in RG as compared to GS has 54,901 citations lesser in RG as compared to GS. Interestingly, there are very few contrary examples too. For example, author 'Tom Maniatis' who has 514 publications with total 371,774 citations in GS have lesser publications (475) but higher citations (385,843) in RG. However, in general, GS is found to be recording higher number of citations for the vast majority of the given set of authors. Further, the difference in citation counts is significantly high between Gs and RG.

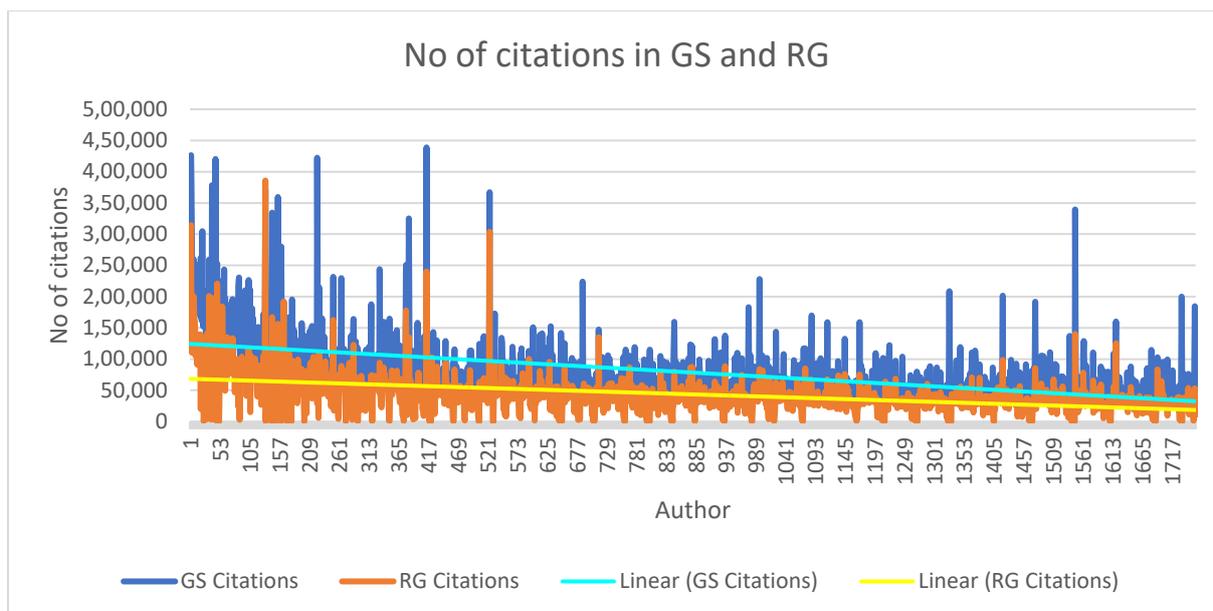

**Figure 4: Citation counts of authors in GS and RG**

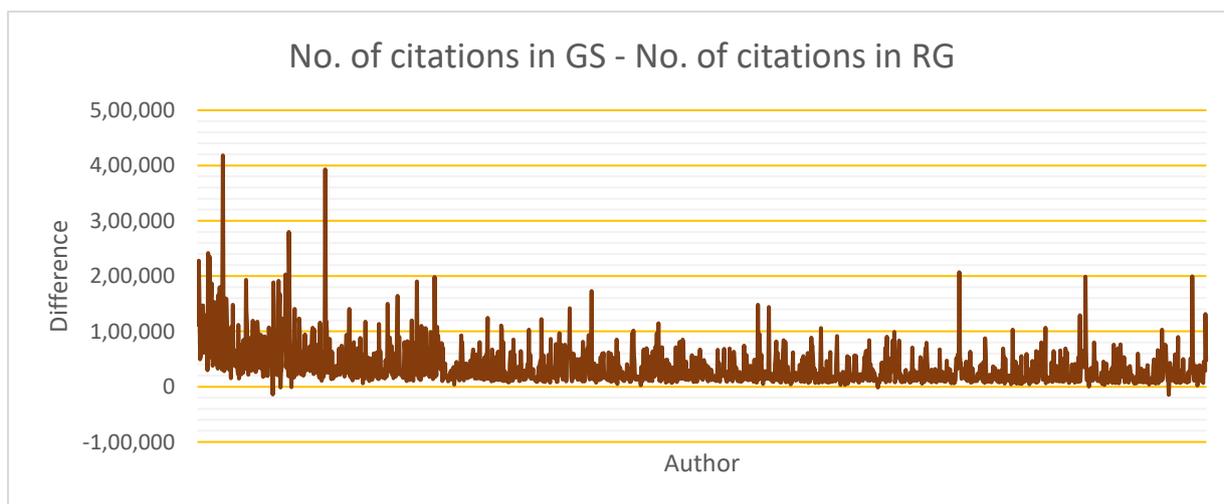

**Figure 5: Difference in citation counts of authors in GS and RG**

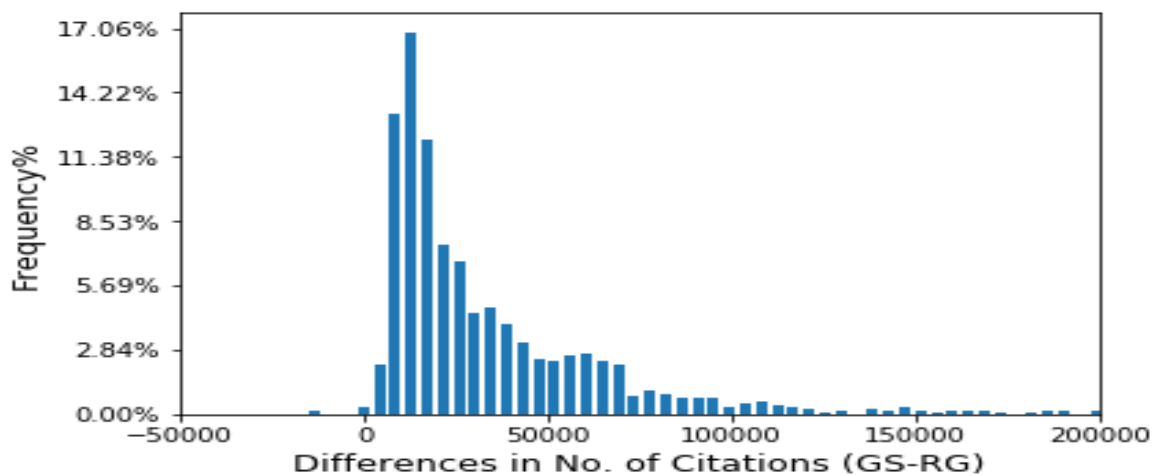

**Figure 6: Frequency of difference in citation counts of authors in GS and RG**

*GS and RG h-index values*

The h-index values for authors in the two platforms is the third parameter compared. For this purpose, we used the h-index values computed by the GS and RG. Both the h-index values include self-citations in their computation. **Figure 7** plots the h-index values for all the authors in the two platforms. It is observed that for majority of the authors the GS h-index value is higher than RG h-index value. **Figure 8** presents the shows the difference value between GS h-index and RG h-index. It is observed that for a vast majority of authors, the difference in GS h-index and RG h-index value is positive. Only 29 authors have slightly higher RG h-index value. For many authors, the difference is even larger than 100. The mean value of the difference is 25. **Figure 9** shows the frequency of difference values in various intervals. It is seen that only about 1.6% authors have GS h-index lesser than RG h-index. For about 60% authors, the difference is in the range 0 – 25. About 38% authors have the difference value greater than 25. The quantum of difference in h-index values, however, is not as large as publication counts and citations. The rank correlation between Gs and RG h-index values is 0.6, as we will see ahead.

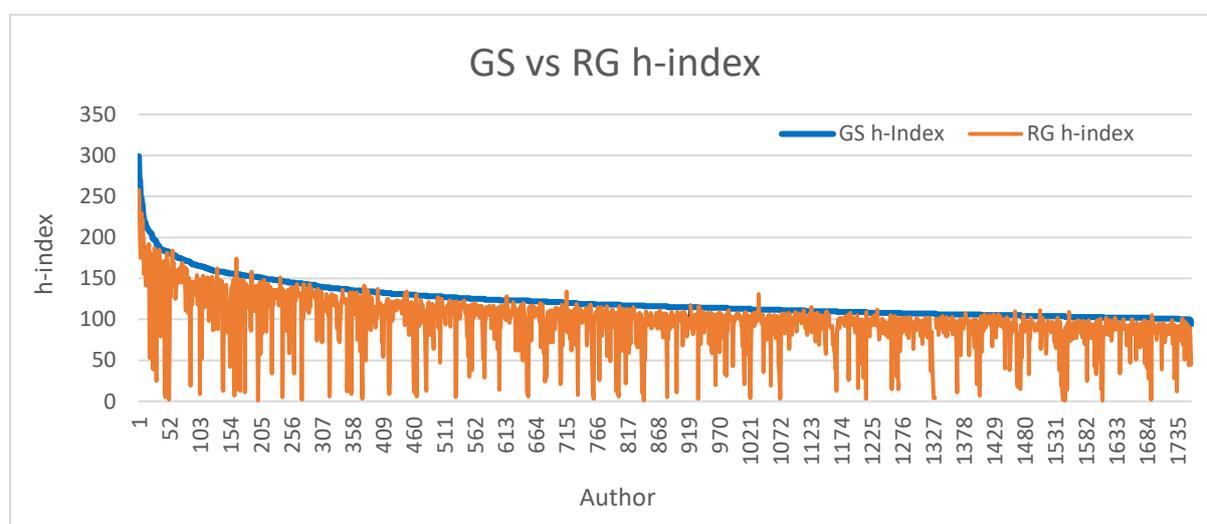

**Figure 7: The h-index values in GS and RG for the 1,758 authors**

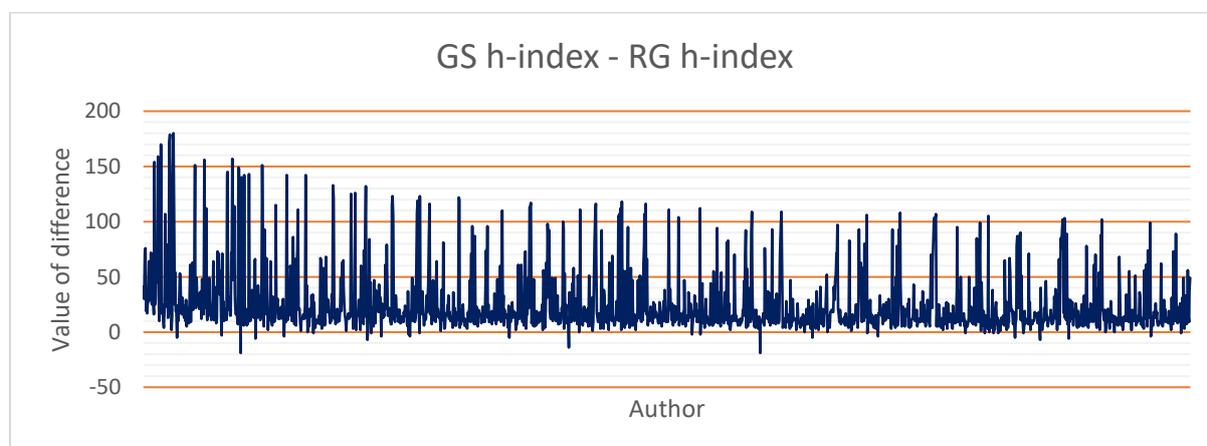

**Figure 8: Difference in GS and RG h-index values of authors**

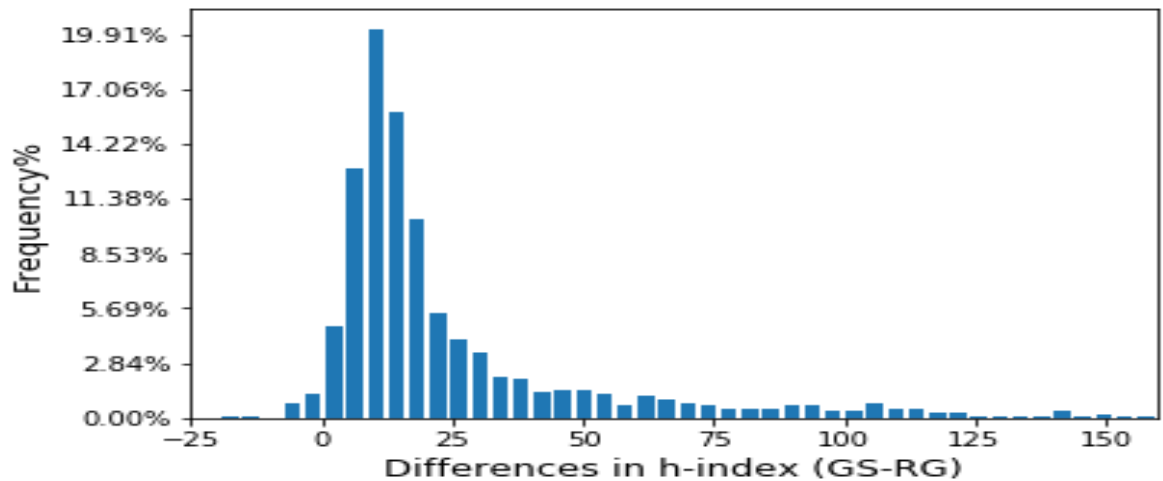

**Figure 9: Frequency in difference in GS and RG h-index values of authors**

*Differences in values of other h-type indices*

Another interesting analysis will be the correlations between other *h*-type indices of GS and RG. Major *h*-type indicators that can be easily estimated are *g*-index (Egghe, 2006) and $\Psi$-index (Lathabai, 2020). These indices are briefly defined as follows:

**g-index** is the highest number *g* of papers that together received $g^2$ or more citations. In other words, *g* is the highest number *g* of papers whose citations average at least to *g* (Tol, 2008, 2009).

**$\Psi$-index** is the highest number $\Psi$ of papers that together received $\frac{\Psi(\Psi+1)}{2}$ or more citations. In other words, $\Psi$ is the highest number $\Psi$ of papers whose citations average at least to $\frac{\Psi+1}{2}$ (Lathabai, 2020).

The need and importance of the estimation arises from the following fact. These *h*-type indices- the *g*-index (Egghe, 2006) and $\Psi$-index (Lathabai, 2020) are not directly available; therefore, it is difficult to conduct exact correlation analysis between RG and GS in terms of these indicators. However, the possibility of estimation of these indicators enables us to conduct correlation analysis that may be equivalent to the correlation analysis between the exact indicators.

These *h*-type indices of scholars for GS and RG can be estimated as follows and can be subjected to correlation analysis.

$$g'(GS) = floor\big(g_{est}(GS)\big) = floor(\sqrt{C_T(GS)}) \quad (1)$$
$$g'(RG) = floor\big(g_{est}(RG)\big) = floor(\sqrt{C_T(RG)}) \quad (2)$$

$$\Psi'(GS) = floor\big(\Psi_{est}(GS)\big) = floor\left(\frac{-1+\sqrt{1+8\,C_T(GS)}}{2}\right)$$
$$= floor(\sqrt{2\,C_T(GS)}) \quad (3)$$

$$\Psi'(RG) = floor\big(\Psi_{est}(RG)\big) = floor(\sqrt{2\,C_T(RG)}) \quad (4)$$

where, $C_T$ denotes the total number of citations in the profile of a scholar. $C_T (GS)$ and $C_T (RG)$ are the total number of citations received by a scholar in GS and RG, respectively. Floor function helps to keep these estimated indices as natural numbers like their actual counterparts.

However, from the analysis conducted in the previous sections, one can notice a significant disparity in the indexing so that information (information found commonly) in RG profile and GS profile of same author can vary significantly. Therefore, it may even result in different scores during performance assessment based on this information. One of the ways is to check whether there is discrepancy in indexing, the determination of different profile types of scholars in both databases can be used. According to Lathabai (2020), an author can have one of the following profile types: regular citation profile, citation excess profile or transition profile.

Regular citation profile is the commonly found profile among authors and it is characterized by $T^2 > C_T$, where $T$ is the total number of publications. In regular citation profile, $C_T$ will be usually less than $T^2$ by a very great margin.

Citation excess profile is rarely found among authors and it is characterized by the presence of one or more heavily cited papers so that $C_T > T^2$.

Transition profile is an intermediate profile of an author which is undergoing a transition from citation excess type to regular citation profile. Though it is usually characterized by $T^2 > C_T$, $T^2$ will not be too high than $C_T$ indicating the chance of a recent transition from a state $C_T > T^2$. Lathabai (2020) gave a method to determine the profile type of an individual. Here, we give a simplified version of the procedure to identify the profile types.

1. Check whether $C_T > T^2$. If it is satisfied, the profile is citation excess profile.
2. If $T^2 > C_T$, compute $\Psi'$).
    (a) If $\Psi' > T$, the profile is a transition profile
    (b) Otherwise, the profile is a regular citation profile

where, $\Psi'$ is the estimated $\Psi$-index that can be obtained using eqns. (3) and (4)

Now, the distribution of scholar profiles in three profile types for both GS and RG is shown in **table 1**.

**Table 1. Distribution of scholar profiles in three profile types for GS and RG and the Jaccard similarity of the distributions**

|  | Regular citation | Citation excess | Transition profile |
|---|---|---|---|
| **Google Scholar (GS)** | 1,685 | 15 | 58 |
| **ResearchGate (RG)** | 1,569 | 81 | 108 |
| **\|GS ∩ RG\|** | 1,545 | 9 | 27 |
| **\|GS ∪ RG\|** | 1,709 | 87 | 139 |
| $J = \frac{\|GS \cap RG\|}{\|GS \cup RG\|}$ | 0.904 | 0.103 | 0.194 |

This may be an indication of the discrepancy in indexing of both the databases. To confirm this discrepancy, Jaccard Similarity index can be used.

From table 1, Jaccard Similarity indices of (i) Citation excess profiles in RG and GS and (ii) Transition profiles in RG and GS, are found to be low. This is a partial indication of the discrepancy in indexing of both the databases. However, as regular citation profiles in both the databases show high value for Jaccard similarity, for confirmation of the discrepancy and also for checking whether the effect of discrepancy is significant among regular citation profile holders, we have to conduct correlation analysis for commonly found authors in the group having regular citation profiles with respect to major productivity indices (estimated) like *g'* and $\Psi'$. **Figures 10** and **11** show the plot of *g'* and $\Psi'$ indices of the 1,545 commonly found scholar profiles of the regular citation type in both GS and RG. The values of co-efficient $\rho$ is found to be 0.6205 and 0.6207 for *g'* and $\Psi'$ indices, respectively. Thus, it is observed that estimated values of the other *h*-type indies show similar level of correlation as that of the actual *h*-index values of the authors in GS and RG.

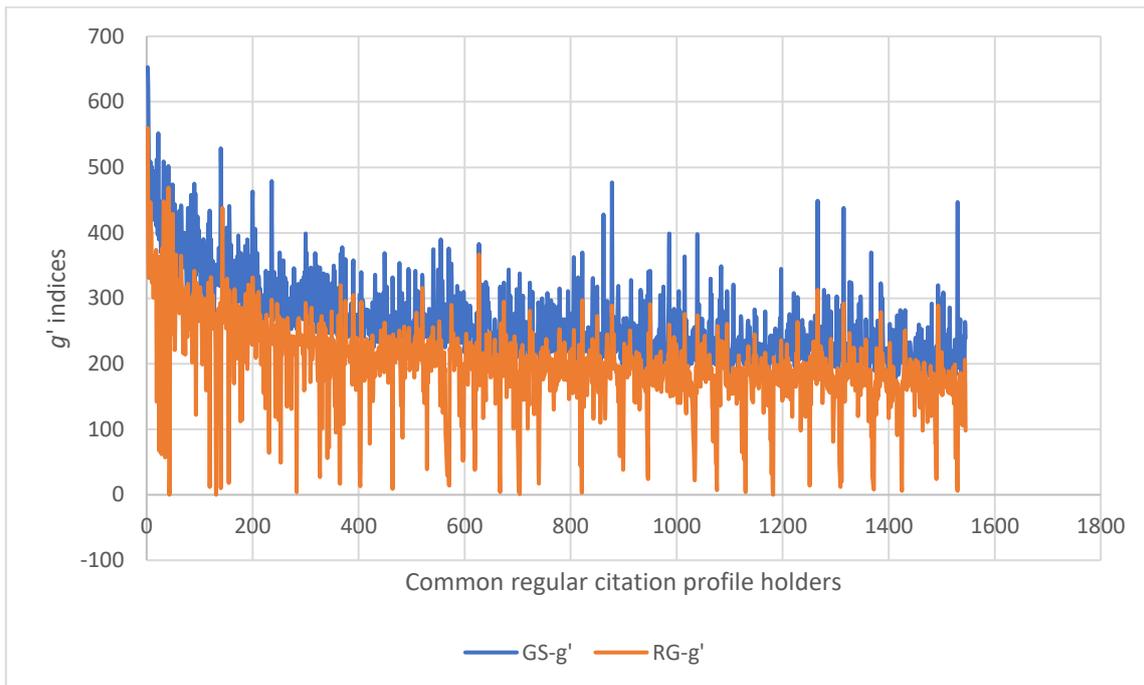

**Figure 10:** *g'* indices of commonly found regular profile holders in GS and RG

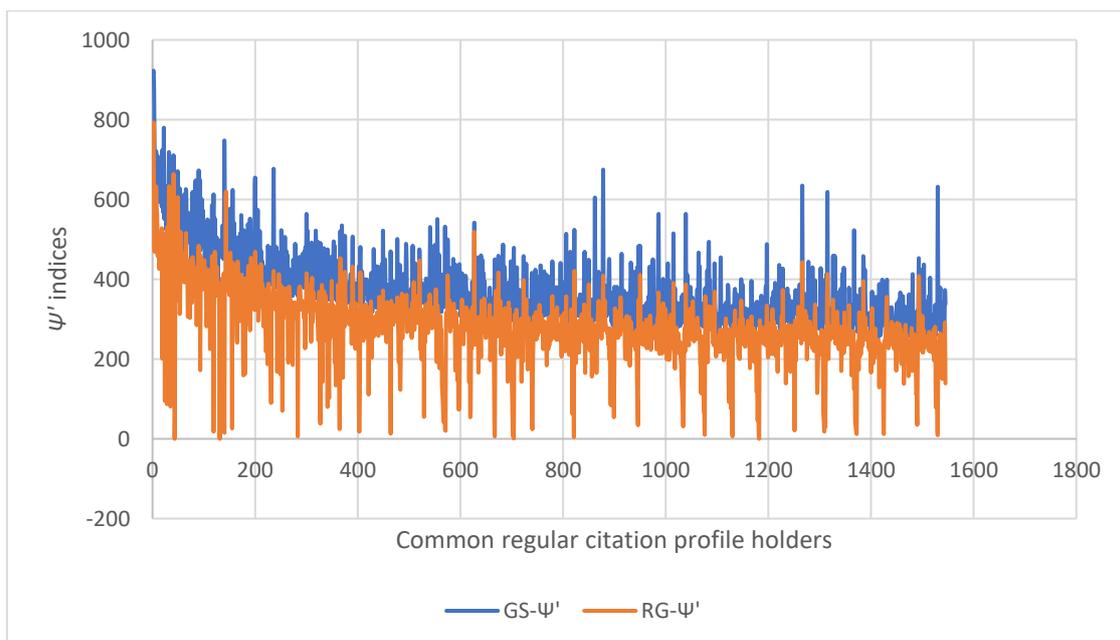

**Figure 11:** $\Psi'$ indices of commonly found regular profile holders in GS and RG

Even though a great share of the regular citation profiles is found commonly in both RG and GS, Spearman's Rank Correlation analysis reveals that there is no strong correlation between the rank/order upon which the scholars can be arranged according to estimated *h*-type productivity indicators. Thus, it can be inferred that discrepancy in indexing can have a significant effect on the productivity/performance assessment of scholars. In other words, the variation in accuracy, velocity and volume of indexing published works and tracking their citations to existing works among databases might end up with significantly difference values for the same productivity/performance indicator.

*RG score compared with GS h-index*

We have also compared the values of GS h-index and RG score of the 1,758 authors. While, GS h-index is a composite score computed from total publications and total citations of an author, the RG score is a composite score having a more complex calculation. Both these scores, however, in a sense represent the scientific reputation of an author. This is why we tried to see whether authors having higher h-index in GS also have higher RG score. **Figure 12** plots the values of GS h-index and the RG score for all the authors. The curve at the top represents GS h-index and the curve at the bottom represent RG score. It is observed there is no clear relationship between GS h-index and RG score values for the authors. While the GS h-index value shows a decreasing pattern from the value of 299 to 100, the RG scores do not show any clear pattern. The RG score is found to vary from lowest value of 2.1 to highest value of 157.37, with an average value of 47.45. For GS h-index, the average value = 122.92, variance = 536.5 and the standard deviation = 23.16. For RG score, the average value = 47.45, variance = 84.98 and the standard deviation = 9.21. Given the fact that RG score is comprised of different components, including the social networking components of the RG, it may be the case that it is not dependent significantly on publications and citations alone. Another interesting thing to note is that the RG score does not show similar differentiation in authors as that of publications, citations and h-index values, as for majority of authors the RG score value is less than 50.

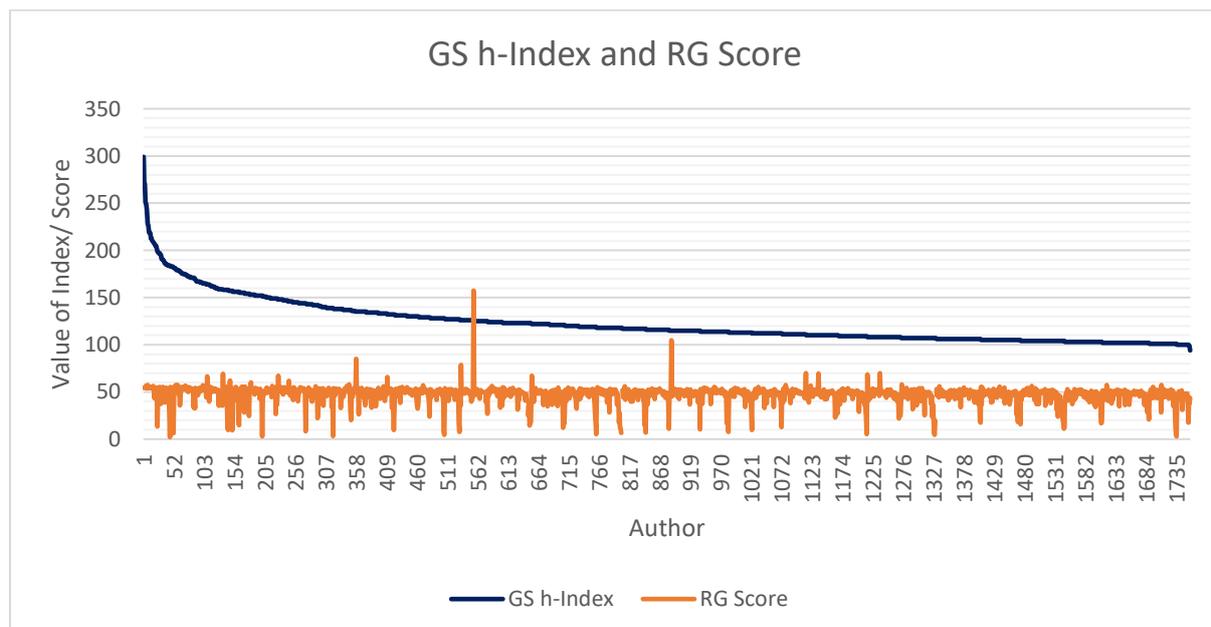

**Figure 12: GS h-index and RG Score for authors**

*Correlations between various counts and metrics in GS and RG*

The Spearman Rank Correlation Coefficients for different counts and metrics of the authors recorded in GS and RG are computed to further understand the relationship between them. **Table 2** shows the different correlation coefficients computed. It can be seen that rank correlation value between GS publications and RG publications is 0.52, indicating moderate positive correlation. Similarly, rank correlation between GS citations and RG citations is 0.55, again indicating moderate positive correlation. GS h-index and RG h-index have a correlation value of 0.6. The RG reads have least correlation with any of the GS values. The RG score shows weak positive correlation with GS publications and GS i-10 index. Its correlation with GS h-index is weaker with a value of 0.31 and quite weak with GS citations with a value of 0.12. Thus, the GS and RG values and metrics show positive correlations of different degrees, but mostly in weak or moderate range.

To understand the computation of RG Score values further, we also computed intra correlations between different values of RG. **Table 3** shows these rank correlation values. It is observed that RG Score has rank correlation value of 0.86 with publications, 0.58 with citations, 0.41 with reads and 0.58 with RG h-index. Thus, the RG Score seems to have highest correlation with number of publications. Given the knowledge that RG Score is computed from five factors: publications, questions, answers, followers and recommendations; it is interesting to observe that RG score has high positive rank correlation with number of publications. The RG h-index shows a correlation value of 0.53 with publications and 0.8 with citations.

**Table 2: Spearman Rank Correlation Coefficients between GS and RG metrics**

| GS values \ RG values | RG Publications | RG Citations | RG h-index | RG Reads | RG Score |
|---|---|---|---|---|---|
| GS Publications | 0.52 | 0.08 | 0.1 | 0.12 | 0.42 |
| GS Citations | 0.10 | 0.55 | 0.38 | 0.21 | 0.12 |
| GS h-index | 0.27 | 0.49 | 0.6 | 0.17 | 0.31 |
| GS i10-index | 0.56 | 0.15 | 0.21 | 0.15 | 0.46 |

**Table 3: Intra-correlation between different values and metrics of RG**

|  | RG Publications | RG Citations | RG Reads | RG Score | RG h-index |
|---|---|---|---|---|---|
| RG Publications | 1 | 0.56 | 0.52 | 0.86 | 0.53 |
| RG Citations | 0.56 | 1 | 0.49 | 0.58 | 0.8 |
| RG Reads | 0.52 | 0.49 | 1 | 0.41 | 0.34 |
| RG Score | 0.86 | 0.58 | 0.41 | 1 | 0.58 |
| RG h-index | 0.53 | 0.8 | 0.34 | 0.58 | 1 |

**Discussion**

The study analysed the publications, citations and different metric values computed by GS and RG, for a large set of highly cited authors. The objective was to understand how similar or dissimilar these counts and metric values are across the two platforms, and if they differ what is the quantum of such differences. Analytical results show that GS has a significant edge in publication and citation counts over RG. The values of various h-type indices for the authors also differ significantly in GS and RG, with GS having a higher value than RG in a vast majority of the cases However, the moderate positive correlations are observed in the values of different h-type indices in the two platforms. Further, the RG reads do not correlate much with any of the GS counts (publications and citations) and metrics (h-index and i-10 index), indicating that reads represent an altogether different dimension. The RG score also shows weak correlations with GS publications and GS i-10 index.

One of the most important and fundamental things to observe is that the publication and citation counts for the same set of authors differ significantly in GS and RG. When complete publication data for all authors is taken together, it is observed that GS records 64% more publications than RG, i.e., GS records 1.64 times more publications than RG. Similarly, the cumulative citation counts for all the authors show that GS records 80% more citations than RG, i.e., 1.8 times more citations than RG. These differences perhaps explain why the h-index and the other h-type indicx values for the authors are higher in GS than RG. Some of the previous studies that analysed GS and RG data also found partly similar results. For example, Thelwall & Kousha (2017b) found that GS always showed more citations for each individual journal than RG. It was also found that RG citations correlated strongly with GS citations. They suggested that correlations are due to the reason that RG is not predominantly tapping a fundamentally different source of data than GS. Martín-Martín, Orduña-Malea & López-Cózar (2018) also found that GS provides more comprehensive citation-related data as compared to other platforms (including RG). Orduna-Malea & Delgado López-Cózar (2017) found that the high-performance authors are not the same across the three studied dimensions of impact (citations, reads, and online mentions) and concluded that lack of correlation in Author Level Metrics from the the platforms might be explained by the fact that each platform offers different documents, targeted to different audiences.

Given the fact that differences in publications and citations and different metrics in GS and RG are clearly seen and the quantum of such differences is significant, it is extremely important to explore the possible reasons for the differences in counts and metric values in the two platforms. We know that RG and GS both extract publication from different databases and their indexing algorithms automatically extract bibliographic data, citations and other information from articles and use it for ranking purposes. Thus, both GS and RG use automated Web crawlers to collect data for publications and citations. Therefore, one needs to know what causes the two platforms to record different publications and citations for the same authors. One may expect that the differences may be due to differences in the crawling algorithms used. However, the exact technical details of the data collection process of the two platforms are not public. In this situation, it is a bit difficult to identify the exact technical reasons that may be causing the difference in publications and citations captured by the two platforms. However, we try to explore some probable reasons for the same by analysing the data used in the study and also by pursuing the very little literature available in this regard.

- The first probable reason for the difference in publications indexed by the two platforms may be the indexing policy of the two platforms. GS has a wider and liberal indexing and coverage policy through which it indexes a wide variety of electronic documents, that have reference and citation links to scholarly articles. GS robots generally try to index every paper from every website they visit, including most major sources and also many lesser-known ones. GS indexes journal and conference papers, theses and dissertations, academic books, pre-prints, abstracts, technical reports and other scholarly literature from all broad areas of research. Scholarly work from a wide variety of academic publishers, professional societies and university repositories, as well as scholarly articles available anywhere across the web are indexed. GS also includes court opinions, inventions and patents. The only exception that GS has is that it excludes documents that are untitled or do not have authors associated. RG on the other hand mainly indexes articles from scholarly sources such as journals, conferences, preprint archives, books etc. It does not include inventions and patents. Therefore, in this sense

- RG is more focused towards indexing scholarly articles from scholarly sources, and has a very good coverage particularly of journal articles.

- The second probable reason for the differences may be the quantum of indexing errors in the two platforms. It has been observed that GS has even included several documents as articles that are merely cover pages, index pages or editor information pages of different journals. Not only these documents are indexed as papers, but they are found added to all author profiles, whose name appear in the document. Further, GS is also found to index papers not belonging to a particular author in his/ her profile. In few cases, GS is strangely found to have indexed a paper to an author profile, because his/ her name appeared in the references section of that paper. This was more so in cases of older papers. RG is also found to have several problems in indexing, but of different kind. First, it is found to have some strange algorithm that if for a given paper it doesn't find a match for the author's profile then it creates some shadow profile for the author and adds the article to that author name, even though the actual author may be having a profile in RG. This results in having duplicate profiles for some authors, both of which do not have complete publication list of the author added. Thus, even though RG may have indexed a paper, it erroneously did not attribute it to the correct author and hence having lesser than actual number of publications for the authors linked to his/ her profile. In our RG crawling process also, we found many cases of same authors having two well-managed profiles with different publication counts.

- The third possible reason for GS recording higher publication counts than RG may be related to the strategy of the two platforms in dealing with the so-called predatory publishing. Marina & Sterligov (2021) indicate that GS is not widely used by research managers, because of its strategy of covering virtually all scientific literature (including predatory publishing) with little quality control. Shamseer et al. (2017) also discussed the GS's lack of immunity from predatory, fake, decisive or shell publications/journals/publishers. Although RG has also been criticized lately for failing to provide safeguards against "ghost journals", publishers with "predatory" publication fees, and fake impact ratings (Memon, 2016), however RG is not as lenient as GS. Therefore, the way GS and RG deal with indexing predatory publishing could be another reason for the difference in publications captured by them.

- The fourth probable reason for the differences may be the process of author attribution to different articles indexed by the two platforms. Van Noorden (2014) points out that RG uses a crawler to find PDF versions of articles on the homepages of authors and publishers and then these are presented as if they had been uploaded to RG by the author. It further states that "*some of the apparent profiles on the site are not owned by real people, but are created automatically – and incompletely – by scraping details of people's affiliations, publication records and PDFs, if available, from around the Web*". Murray (2014) also found that in 2014 a dormant account on RG, using default settings, was automatically attributed to more than 430 publications. Martín-Martín, Orduna-Malea and López-Cózar (2016) while highlighting the advantages and disadvantages of GS, RG, Mendeley etc. states that RG has no automatic update mechanism (and that a co-author must upload the article document) whereas GS has automatic updates. Thus, while GS has a more automated mechanism of author attribution to a publication, RG sometimes fail to automatically attribute publications to the correct author. Therefore, it appears that in several cases, author confirmation becomes a prerequisite for adding a publication in that author's RG profile. Thus, absence of confirmation may result in some papers not been added to RG profile of authors. And at the same time, absence of

> such confirmation in GS may result in GS erroneously adding papers not belonging to the author in his/ her profile.

The differences in publication indexing and their author attribution in the two platforms indirectly result in differences in citations captured by the two platforms. A more detailed analysis of the exact technical details of the data collection process of GS and RG, however, is required in order to concretely identify the exact reasons for differences in their publication counts and citations for the same authors.

Another interesting thing observed in the paper is with respect to the RG score and its correlation with various parameters in GS as well as RG. It is observed that RG score is very weakly correlated with GS publications, citations and h-index (see table 2). Of particular interest is also the variation in RG score and GS h-index of the various authors (see figure 12). It is observed that while GS h-index has a clearly visible variation (average value = 122.92, variance = 536.5 and standard deviation =23.16), the RG score value doesn't vary much from one end to another and rather remains centred around a particular value range (average value = 47.45, variance = 84.98 and standard deviation = 9.21). For example, the RG score value for the author having GS h-index 299 is 54.28 and for the author having GS h-index value 100, it is 51.63. Thus, RG score appears to be lacking the differentiating ability between different authors and may not be an appropriate measure of scientific reputation of an author (usually measured in terms of publications and citations), as also pointed out in several previous studies (Kraker & Lex, 2015; Jordan, 2015, Yu et al., 2016; Orduna-Malea et al., 2017; Lepori, Thelwall & Hoorani, 2018; Copiello & Bonifaci, 2018; Copiello & Bonifaci, 2019; Copiello, 2019).

One may be interested to know as to how RG score is computed in order to further understand the observed patterns. The details available in RG state that computation of RG score is based on five factors: *publications in indexed journals* (if the journal is not indexed, the RG score will not change substantially), *questions* (this is given more weightage than all other factors), *answers*, *followers* (least weightage) and *recommendations* (if a RG user recommends research items, questions and answers, it will boost RG score a lot). RG further states that the "*RG score measures scientific reputation based on how your work is received by your peers*"[7], indicating that it focuses on how effectively a scientist is able to communicate to the community. RG help page also states that "*RG Score is calculated based on any contribution you share on ResearchGate or add to your profile, such as published articles, unpublished research, projects, questions, and answers*". It further adds that "*Our algorithm looks at how your peers receive and evaluate these contributions, and who they are*". This suggests that "*The higher the RG Scores of those who interact with your research, the more your own score will increase. A low-quality contribution probably won't attract positive feedback and recognition from the community, so it won't contribute to your score in any significant way*".

The intra-correlations in RG metrics (see table 3) confirm the above as it is observed that RG score correlates highest with RG publications but has only moderate correlation with citations, RG h-index and reads. Therefore, it appears that the social networking feature of RG (as seen in the questions, answers, followers and recommendations) also contributes significantly to the RG score. Therefore, RG score may be a more useful measure of "*how a scientist is able to communicate with peers*" rather than a measure of scientific reputation in the traditional sense, as also pointed out by the previous studies (Kraker & Lex, 2015; Jordan, 2015; Orduna-Malea et al., 2017; Copiello & Bonifaci, 2018; Copiello & Bonifaci, 2019). Further, the correlations of RG score with GS metrics show that it represents a different aspect of scientific publishing and communication rather than the traditional publication-citation based traditional impact assessment measures.

---

[7] https://explore.researchgate.net/display/support/RG+Score

**Conclusion**

The article explored the publications, citations and different metrics in ResearchGate and Google Scholar for a set of highly cited authors. It has been found that Google Scholar indexes more publications and captures more citations as compared to ResearchGate for a vast majority of authors. The difference in publications and citations captured by Google Scholar and ResearchGate is found to be quite high. The correlations between different metrics in Google Scholar and ResearchGate are also computed and it is found that h-index values of Google Scholar and ResearchGate show moderate positive correlation. However, the RG score doesn't correlate well with Google Scholar h-index. Similarly, the RG reads also do not correlate with different Google Scholar metrics. The coverage policy, indexing errors, author attribution mechanism and the strategy to deal with predatory publishing are found to be the main probable reasons for the observed differences in the two platforms.

A major effect of the difference in indexing and tracking mechanisms of ResearchGate and Google Scholar is that, for the same individual scholar profile, the type of profile might be different. This will definitely affect the productivity assessment using these profiles. Even if the profile type is the same, it is observed that the effect of difference in indexing and tracking is so intense that the resulting productivity measures will be significantly different, posing a challenge to the reliability of assessment using these databases. The different metrics computed by Google Scholar (such as h-index) and ResearchGate (such as RG score) also show different behaviours, indicating that they may be measuring different dimensions of the authors research profile. Therefore, one has to be careful to use and interpret these metrics for assessment of scientific reputation of authors. More studies on differences in Google Scholar and ResearchGate at the level of institutions and/ or journals can be done in order to have a better understanding of the similarities and differences in the Google Scholar and ResearchGate platforms.

**Acknowledgements**

This work is partly supported by the extramural research grant no: MTR/2020/000625 from Science and Engineering Research Board, India to the first author.